\documentclass[10pt,conference]{IEEEtran}
\AtBeginDocument{%
 \providecommand\BibTeX{{%
   \normalfont B\kern-0.5em{\scshape i\kern-0.25em b}\kern-0.8em\TeX}}}

\usepackage{xspace}
\newcommand{\ie}{\textit{i.e., }}
\newcommand{\eg}{\textit{e.g., }}
\newcommand{\etal}{\textit{et al. \xspace}}
\newcommand{\taxonomy}{25\xspace}
\usepackage{longtable}
\usepackage{subcaption}
\usepackage{alphalph}
\usepackage{amsmath,amsfonts}
\usepackage{algorithmic}
\usepackage{graphicx}
\usepackage{comment}
\usepackage{textcomp}
\usepackage{xcolor}
\usepackage{adjustbox}
\usepackage{textcomp}
\usepackage[numbered]{bookmark}
\usepackage{soul}
\usepackage{booktabs}
\usepackage{multirow}
\usepackage{float}
\usepackage[many]{tcolorbox}
\usepackage{fontawesome}
\usepackage{graphicx}
\usepackage{url}
\usepackage{tabularx} 
\usepackage[justification=centering]{caption}
\usepackage[switch]{lineno}
\usepackage{ragged2e}
\usepackage{dirtytalk}
\usepackage{multicol}
\usepackage{csquotes}
\usepackage{pgfplots, pgfplotstable}
\usepackage{pgf-pie}
\usepackage{colortbl}
\usepackage{xspace}
\usepackage{array}
\usepackage{listings}
\usepackage{balance}
\usepackage{cite}
\usepackage{supertabular} 
\newcolumntype{L}{>{\arraybackslash}m{16cm}}
\usepackage[flushleft]{threeparttable}
\usepackage{supertabular}
\usepackage{mdframed}
\lstdefinestyle{mystyle}{
    basicstyle=\ttfamily\scriptsize, 
    keywordstyle=\color{blue},
    commentstyle=\color{gray},
    stringstyle=\color{red},
    numbers=left,
    numberstyle=\tiny\color{gray},
    numbers=none, 
    stepnumber=1,
    numbersep=4pt,
    breaklines=true,
    breakatwhitespace=true,
    columns=flexible,
    frame=single,
    captionpos=b,
    escapeinside={@}{@} 
}
  \pgfplotstableread[row sep=\\,col sep=&]{
        interval & carT & sd\\
     }\mydata

\usetikzlibrary{shapes.geometric, arrows} 

\tikzstyle{startstop} = [rectangle, rounded corners, minimum width=3cm, minimum height=1cm,text centered, draw=black, fill=black!10]

\tikzstyle{io} = [trapezium, trapezium left angle=70, trapezium right angle=110, minimum width=3cm, minimum height=1cm, text centered,text width=2cm, draw=black]

\tikzstyle{process} = [rectangle, minimum width=3cm, minimum height=1cm, text centered, text width=3cm, draw=black]

\tikzstyle{decision} = [diamond, minimum width=3cm, minimum height=1cm, text centered,text width=2cm, draw=black]

\tikzstyle{arrow} = [thick,->,>=stealth]

\tikzstyle{database} = [cylinder, shape border rotate=90, draw=black,minimum height=2cm,minimum width=3cm, text centered,text width=0.6cm]

\newlength\q 

        {\begin{itemize}\setlength{\itemsep}{0pt}\setlength{\parsep}{0pt}
	\setlength{\topsep}{0pt}\setlength{\parskip}{0pt}}
	{\end{itemize}}
	{\begin{enumerate}\setlength{\itemsep}{0pt}\setlength{\parsep}{0pt}
	\setlength{\topsep}{0pt}\setlength{\parskip}{0pt}}
	{\end{enumerate}}
	{\begin{enumerate}\setlength{\itemsep}{0pt}\setlength{\parsep}{0pt}
	\setlength{\topsep}{0pt}\setlength{\parskip}{0pt}\setlength{\leftmargin}{0pt}}
	{\end{enumerate}}   

\tcbset{
    sharp corners,
    before skip = 0.2cm,    
    after skip = 0.5cm      
} 

\newtcolorbox{boxK}{
    sharpish corners, 
    boxrule = 0pt,
    toprule = 4.5pt, 
    enhanced,
    fuzzy shadow = {0pt}{-2pt}{-0.5pt}{0.5pt}{black!35} 
}
\newtcolorbox{boxA}{
fontupper = \small, 
    boxrule = 1.5pt,
    colframe = black 
}

\newcommand{\ali}[1]{\textcolor{green}{{\it [Ali says: #1]}}}

\begin{document}

\author{
\IEEEauthorblockN{Eman Abdullah AlOmar\IEEEauthorrefmark{1},
Luo Xu\IEEEauthorrefmark{1},
Sofia Martinez\IEEEauthorrefmark{1},
Anthony Peruma\IEEEauthorrefmark{2},
Mohamed Wiem Mkaouer\IEEEauthorrefmark{3}, \\
Christian D. Newman\IEEEauthorrefmark{4},
Ali Ouni\IEEEauthorrefmark{5}}
\IEEEauthorblockA{\IEEEauthorrefmark{1}Stevens Institute of Technology, Hoboken, NJ, USA\\
\IEEEauthorrefmark{2}University of Hawaii at Manoa, Honolulu, HI, USA\\
\IEEEauthorrefmark{3}University of Michigan-Flint, Flint, MI, USA\\
\IEEEauthorrefmark{4}Rochester Institute of Technology, Rochester, NY, USA\\
\IEEEauthorrefmark{5}ETS Montreal, University of Quebec, Montreal, QC, Canada\\ 
\text{\{ealomar,lxu41,smartine1\}@stevens.edu}, 
\text{peruma@hawaii.edu},
\text{mmkaouer@umich.edu},
\text{cdnvse@rit.edu},
\text{ali.ouni@etsmtl.ca}\\
}}

\title{\huge ChatGPT for Code Refactoring: Analyzing Topics, Interaction, and Effective Prompts} 
\maketitle
\begin{abstract}
\textcolor{black}{Large Language Models (LLMs), such as ChatGPT, have become widely popular and widely used in various software engineering tasks such as refactoring, testing, code review, and program comprehension. Although recent studies have examined the effectiveness of LLMs in recommending and suggesting refactoring, there is a limited understanding of how developers express their refactoring needs when interacting with ChatGPT}. In this paper, our goal is to explore interactions related to refactoring between developers and ChatGPT to better understand how developers identify areas for improvement in code, and how ChatGPT addresses developers' needs. \textcolor{black}{Our approach involves text mining 715 refactoring-related interactions from 29,778 ChatGPT prompts and responses, as well as the analysis of developers' explicit refactoring intentions}. Our results reveal that 
 \textcolor{black}{(1) refactoring interactions between developers and ChatGPT encompass \taxonomy themes including `Quality', `Objective', `Testing', and `Design', 
  (2) ChatGPT's use of affirmation phrases such as `certainly' regarding refactoring decisions, and apology phrases such as `apologize' when resolving refactoring challenges, and (3) our refactoring prompt template enables developers to obtain concise, accurate, and satisfactory responses with minimal interactions}.  We envision our results enhancing researchers and practitioners 
  understanding of how developers interact with LLMs 
   during code refactoring. 
\end{abstract}

\section{Introduction}
\label{Section:Introduction}

\textcolor{black}{The recent advances in Artificial Intelligence (AI) are revolutionizing computing education in general and software engineering in particular. In fact, the ability of Large Language Models (LLMs) to harness massive amounts of multimodal information has enabled them to perform a variety of tasks that are known to depend on human intervention \cite{nathalia2023artificial,ahmad2023towards}. Leveraging information from open-source software (OSS) repositories, these LLMs are emerging as assistive technologies for developers in various software engineering 
 tasks, such as code search \cite{wangayou}, code quality \cite{white2023prompt}, repair \cite{haque2023potential}, program comprehension \cite{ma2023scope},  generation \cite{feng2023investigating}, completion \cite{pudari2023copilot}, and translation \cite{jiao2023chatgpt}. The promising results of LLMs, in general, and ChatGPT, in particular, have grown in popularity quickly within the software engineering community. A recent GitHub survey with 500 US-based developers \cite{futurism}, shows that up to 92\% have AI support integrated into their development environments, and 70\% reported an improvement in their coding productivity. Similarly, recent research highlights the outstanding performance of these models when tested against traditional solutions using existing benchmarks \cite{sun2023automatic}.}

\begin{figure*}[th]
\centering 
\includegraphics[width=0.8\textwidth]{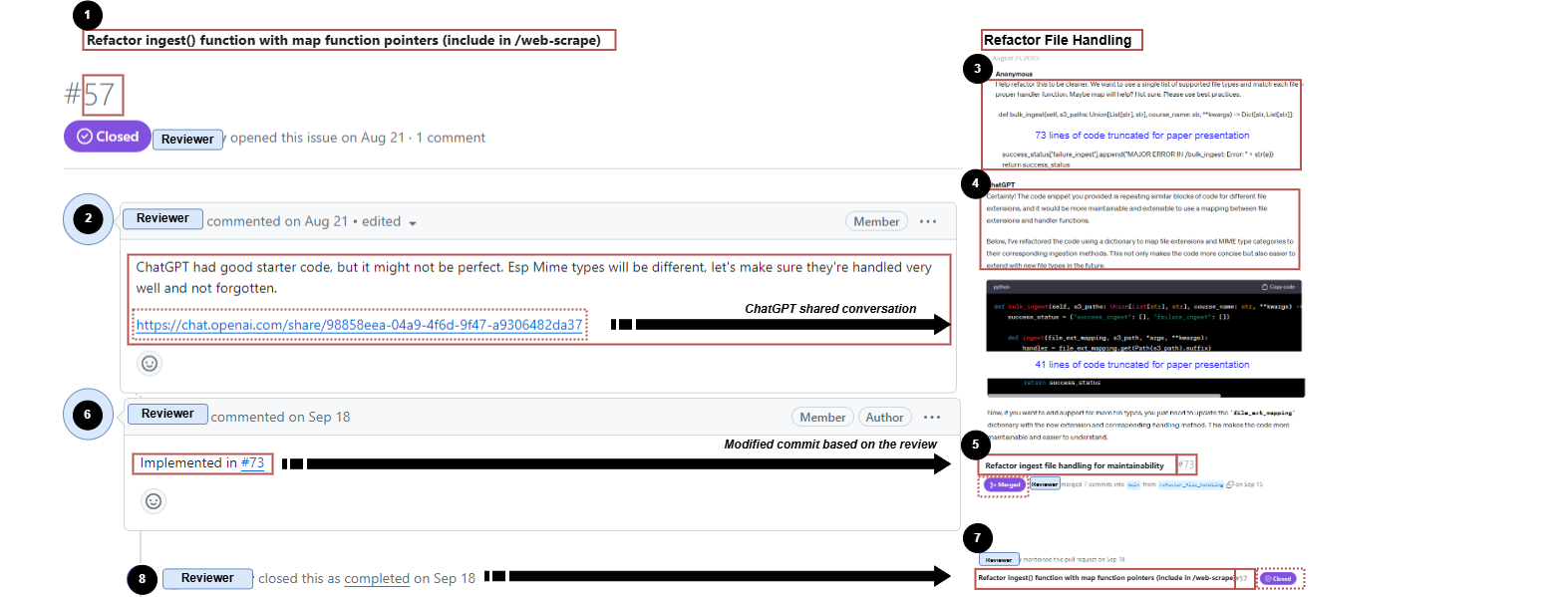}
\caption{An illustration of how a developer interacts with ChatGPT and how the generated outcomes have been included in a pull request that was accepted and deployed in production \cite{Example}.}
\label{fig:example}
\vspace{-.4cm}
\end{figure*}

\textcolor{black}{Despite the existence of built-in models in Integrated Development Environments (IDEs), many developers use conversational models, which are designed to receive text-based requests (\ie prompts) and generate human-like output texts. The interactive nature of such models has increased their popularity among developers, as they can respond to a broader range of queries besides just code. Built-in models, such as GitHub Copilot\footnote{\url{https://github.com/features/copilot}}, provide laser-focused recommendations on current coding tasks, but conversational models, such as ChatGPT, can assist with a wider range of tasks. Therefore, recent studies have shifted from evaluating the capability of ChatGPT, to analyzing collaborative practice and prompt engineering, to improve how developers can foster adequate information that matches their expectations \cite{hao2024empirical,alomar2024refactor}.} 


While LLMs have been heavily solicited for a variety of tasks, little is known about their response to prompts related to code refactoring. Refactoring, by definition, involves enhancing the internal structure of code without changing its external behavior \cite{Fowler:1999:RID:311424,mens2004survey}. Due to the subjective nature of refactoring, where several equally valid solutions might exist for a given scenario \cite{Fowler:1999:RID:311424}, it is intriguing to observe how language models handle refactoring requests and which quality attributes they prioritize during code optimization.
\textcolor{black}{Although there are emerging studies analyzing the refactoring capability of ChatGPT \cite{pomian2024next,liu2025exploring}, they are mainly focused on benchmarking ChatGPT's ability to correctly recommend refactorings.  
Consequently, it is essential to conduct a more detailed examination to comprehend developers' expectations regarding AI-aided refactoring. Additionally, prior research has not explored situations where developers openly show disappointment with the model's output. }

Figure \ref{fig:example} presents an illustration regarding an issue filed to revise the \texttt{ingest()} method. In response to a prompt,  
  ChatGPT delivered an updated code version, clearly stating that the refactoring aimed to enhance \textit{maintainability}. This purpose is also reflected in the concluding title of the pull request that eventually integrated the changes into the live codebase. From this instance, it is evident that ChatGPT targets \textit{maintainability} as a quality attribute in its code optimization process, with the model confirming its quality improvements by using the word `certainly.' Thus, we are motivated to further extract and analyze additional quality attributes to gain a deeper understanding of the refactoring strategies utilized by ChatGPT.

\textcolor{black}{The goal of this paper is to explore what developers care about when they request ChatGPT to refactor their code. Specifically, we aim to identify, categorize, and analyze the patterns in developers' prompts, to extract the main criteria developers rely on to develop a decision about accepting or rejecting the model's response. Furthermore, we synthesize this extracted knowledge to craft a prompt template that can accurately reflect developers' expectations. Our study is based on analyzing 715 refactoring-related interactions collected from 29,778 publicly shared conversations between developers and ChatGPT. These conversations can be found attached to their posted GitHub commits, issues, and files using ChatGPT sharable links. The existence of such traces of interactions is evidence of how developers are trending toward using model output to address their problems, and for possible code alternatives. According to our goal, we aim to answer the following research questions:}

\begin{itemize}
\item \textbf{RQ$_1$:} \textit{What conversation topics ChatGPT takes into consideration when generating responses to refactoring requests?} To answer this research question, we aim to extract all the coding contexts in which refactoring was performed, whether directly requested by the developer or proposed by ChatGPT as a response. To do so, we conducted a thematic analysis of 715 interactions to design a taxonomy of refactoring contexts. The taxonomy contains 4 root themes, with a total of 25 sub-themes. 

\item \textbf{RQ$_2$:} \textit{In which scenarios does ChatGPT provide unsatisfactory responses during the refactoring conversations?} 
 In this research question, we are interested in exploring which scenarios ChatGPT's response was found to be confusing or unsatisfactory. Therefore, we analyze the model's responses that tend to include an apology or an affirmation statement during the refactoring interactions. 

\item \textbf{RQ$_3$:} \textit{\textcolor{black}{What is the performance of the proposed prompt, compared to the dataset?}} We address this research question by designing a prompt template that would efficiently refactor the input code. We evaluate the efficiency of the template by assessing its ability to reproduce in \textit{one shot} the exact refactored code that took developers several rounds of prompts to reach it. We also measure our prompt's length and compare it with the length of the conversations to see if it can lead to the same result with fewer tokens.  
\end{itemize}

Our dataset and artifacts are available for replication and extension purposes \cite{ReplicationPackage}.

\section{Study Design}
\label{Section:Methodology}

\subsection{Data Collection and Curation}

This research employs the DevGPT dataset \cite{xiao2024devgpt}, which comprises extensive data on open-source projects, including code files, commits, issues, Hacker News, pull requests, and discussions. Excluding Hacker News, these elements are all hosted and integrated on GitHub. To collect data from multiple sources, we utilize the following approach:

\textbf{Step \#1: Data Collection:} Initially, the data was collected from the DevGPT dataset \cite{xiao2024devgpt}. This dataset is composed of various JSON files structured into snapshots.

\textbf{Step \#2: Data Extraction:} JSON files were retrieved and organized into distinct categories based on their source type.  \textcolor{black}{A summary of the extracted data is as follows: Shared ChatGPT links (4,733), GitHub or Hacker News references (3,559), ChatGPT prompts and response (29,778), Code snippets (19,106), Refactoring GitHub commits (470), Refactoring GitHub issues (69), and Refactoring GitHub code files (176).}

\textbf{Step \#3: Data Transformation:} The JSON files underwent processing to be converted into a structured, relational table format before being imported into the database.

\textbf{Step \#4: Data Translation:} Given that some interactions were in non-English languages, we used the Google Translate library to assist in converting non-English content into English. Such multilingual analysis enhances the clarity and accessibility of the dataset. Once the translation is complete, the database is updated, in a methodical manner, with the new translated text.

\textbf{Step \#5: Data Preprocessing:} Titles, bodies, prompts, and responses from different sources underwent cleaning, had stop-words removed, and were tokenized.

We developed a pipeline to perform these tasks, starting with the JSON files as input and producing the desired subset as a database output. Our pipeline integrates multiple technologies, including \textit{SQLite} for handling data, \textit{FastText} to identify conversations in languages other than English, the \textit{Google Translator} library for translating these dialogues, and \textit{NLTK \& Spacy} for detailed cleaning and tokenization. Additionally, \textit{Dask} is used to enhance efficiency through concurrent processing during both cleaning and tokenization stages. 

\subsection{Taxonomy Building and Refinement} \label{thematicanalysis} 
 As our research focuses on refactorings, our examination is confined to sources where refactorings were discussed in developer-ChatGPT exchanges. Initially, we extracted all conversations from the original dataset. To confirm that the software artifacts concerned refactoring, we concentrated on prompts illustrating developers' intent to apply refactoring (\textit{i.e.}, containing the keyword `\textit{refactor}'). Searching for the term `\textit{refactor}' is a common method for identifying refactorings in natural language text, and is used by numerous related works to identify refactoring-related commits and text \cite{murphy2008gathering,alomar2019can,di2018preliminary,Ratzinger:2008:RRS:1370750.1370759,alomar2021toward,alomar2021icse}. This process culminated in selecting three types of sources: commits, issues, and files. Ultimately, to minimize false positives, our analysis was restricted to the presence of \textit{`refactor'} in the prompt for each source type. We found a few data points that were duplicates or contained invalid links; we excluded all such instances. The initial processing stage removed 13 duplicated conversations from the dataset, alongside 53 invalid ChatGPT links.  The procedure led to the examination of 470 commits, 69 issues, and 176 files. 

\textcolor{black}{The purpose of the manual analysis is to classify the discussion in developer-ChatGPT interactions. Due to the absence of established taxonomies for tasks associated with refactoring through ChatGPT, we utilized a thematic analysis approach following the guidelines of Cruzes \etal \cite{cruzes2011recommended} when examining ChatGPT responses. Thematic analysis is widely regarded in the Software Engineering literature (\eg \cite{Silva:2016:WWR:2950290.2950305,calefato2023lot}) as a method for detecting and documenting patterns (or \say{themes}) within a set of descriptive labels, referred to as \say{codes}. For each refactoring response from ChatGPT, the analysis was conducted through the following steps: 
i) Initial reading of the responses; ii) Creating preliminary codes (\ie labels) for each response; iii) Translating codes into themes, sub-themes, and more comprehensive themes; iv) Refining themes to identify possibilities for consolidation; v) Defining and naming the final themes while developing a model of higher-order themes supported by evidence.}

The procedures outlined earlier were carried out separately by two of the authors. \textcolor{black}{One author independently labeled the ChatGPT responses, while the other focused on examining the taxonomy draft. After each cycle, both authors convened to refine the taxonomy.} For the manual coding process, we employed a spreadsheet tool with tagging functions. This spreadsheet offered the annotators the following details: (1) the prompt given to ChatGPT, (2) ChatGPT’s response, and (3) a link for ChatGPT file sharing. During the study, \textcolor{black}{one author} possessed 7 years of research experience in refactoring, whereas \textcolor{black}{the other author} had 11 years of experience in the same field.

\textcolor{black}{It is important to note that the approach is not a single-step process, with each round focusing on approximately one-third of the inspection instances. Once the two authors have inspected all 715 instances, we solved the conflicts in 6\% of the cases. Conflicts can be attributed to two design choices. Firstly, the absence of predefined categories was a key factor. This meant that discrepancies arose when two authors independently assigned semantically equivalent, yet distinct, labels to describe a given automated task. Secondly, our cautious approach to defining conflicts contributed to this observation. We define an instance as a conflict if two authors assign different sets of labels, even if there is partial overlap between the two sets. Throughout the analysis of the codes, many first-cycle codes were integrated into other codes, renamed, or completely eliminated. During the process when the authors worked to convert the information into themes, they had to shuffle, fine-tune, and occasionally reclassify data into either different or new codes. For instance, we consolidated the initial categories such as \say{\textit{incorrect refactoring}}, \say{\textit{behavior preservation violation}}, \say{\textit{separation of other changes from refactoring}}, and \say{\textit{interleaving other changes with refactoring}} into the overarching category of \say{\textit{Objective}}.} \textcolor{black}{This method resulted in the development of \taxonomy themes, which were then used to create a hierarchical taxonomy that covers refactoring tasks found in ChatGPT responses.}

\subsection{\textcolor{black}{Refactoring Prompt Template Construction  and Evaluation}} \label{promotconstruction}  \textbf{Step \#1: Prompt Construction.} In interactions with ChatGPT, developers articulate software development concepts using natural language. Due to the varied methods developers use to describe issues, it is impractical to rely solely on automated methods for analyzing prompts and responses. \textcolor{black}{Thus, we conducted a manual examination to construct a prompt for refactoring in three rounds. Two authors independently reviewed the developer-ChatGPT interactions (\ie prompt, response, and code of each interactions), constructed and tested the prompt, resolving any discrepancies through discussion. In the first round, two authors equally divided the ChatGPT-developer conversations.  Then, we started by manually reviewing the unique set of prompts in which developers initiated a conversation with ChatGPT about refactoring. Based on our observations, we recognized that there is a certain format that leads to a shorter interaction between ChatGPT and developers. The effective format usually involves a precise breakdown of prompt explain to ChatGPT what role it is playing, the working set that the developer is using, the refactoring task, the project specifications, requirements, output format example, the operating system being used, the installed tools, and how it is being run.  
 Consequently, we construct our first prompt. Then, in the second round, we review the literature on the realm of LLM refactoring \cite{gehring2023deterministic,white2024chatgpt,shirafuji2023refactoring,depalma2024exploring,alomar2024refactor,pomian2024next,gao2024context,choi2024iterative,wu2024ismell,ishizue2024improved,gautam2024refactorbench,cui2024one,cui2024three,zhang2025move,gao2024preference,zhang2024copilot,zhang2024refactoring} to better understand what prompts have been considered successful. We found that this is a multi-faceted topic, with research expanding into different LLMs, different tools, different prompting techniques, and different refactoring types. From these studies, we noticed that there are certain prompting techniques that are better for refactoring. Thus, we implemented these techniques into the prompt by refining our drafted prompt to include any missing attributes accordingly in our third iteration. The authors evenly split the instances for review. If there were discrepancies, the authors would engage in extensive discussions until a consensus was reached. These discussions often revolved around whether there is noteworthy information to be added in our prompt. Ultimately, both authors agreed on the constructed prompt after discussing each instance.} 

 \textcolor{black}{\textbf{Step \#2: Prompt Evaluation.} After constructing the prompt, we evaluated it by reproducing the corresponding developer's prompt from DevGPT, which serves as our ground truth. The evaluation was conducted using three key metrics: number of turns, prompt length, and response length. To analyze the data, we first tested normality using the Shapiro-Wilk test and found that the distribution of the metrics did not follow a normal distribution. As a result, we applied the Mann-Whitney U test \cite{conover1998practical}, a non-parametric statistical test, to compare the two independent groups. The null hypothesis is defined by the absence of variation in the metric values of the developer's prompt and our prompt. Thus, the alternative hypothesis indicates that there is a variation in the metric values. Furthermore, the variation between values of both sets is considered significant if its associated \textit{p}-value is less than 0.05. Furthermore, we use the Cliff's Delta ($\delta$) \cite{cliff1993dominance}, a non-parametric effect size measure, to estimate the magnitude of the differences between DevGPT developer prompt and our prompt. Regarding its interpretation, we follow the guidelines reported by Romano \etal \cite{romano2006appropriate}: Negligible for $\mid \delta \mid< 0.147$, Small for $0.147 \leq \mid \delta \mid < 0.33$, Medium for $0.33 \leq \mid \delta \mid < 0.474$, and Large for $\mid \delta \mid \geq 0.474$.}

\section{Experimental Results}
\label{Section:Result}

\subsection{RQ$_1$: \textcolor{black}{What conversation topics ChatGPT takes into consideration when generating responses to refactoring requests?}}

\begin{figure*}[t]
\centering 
\includegraphics[width=0.8\textwidth]{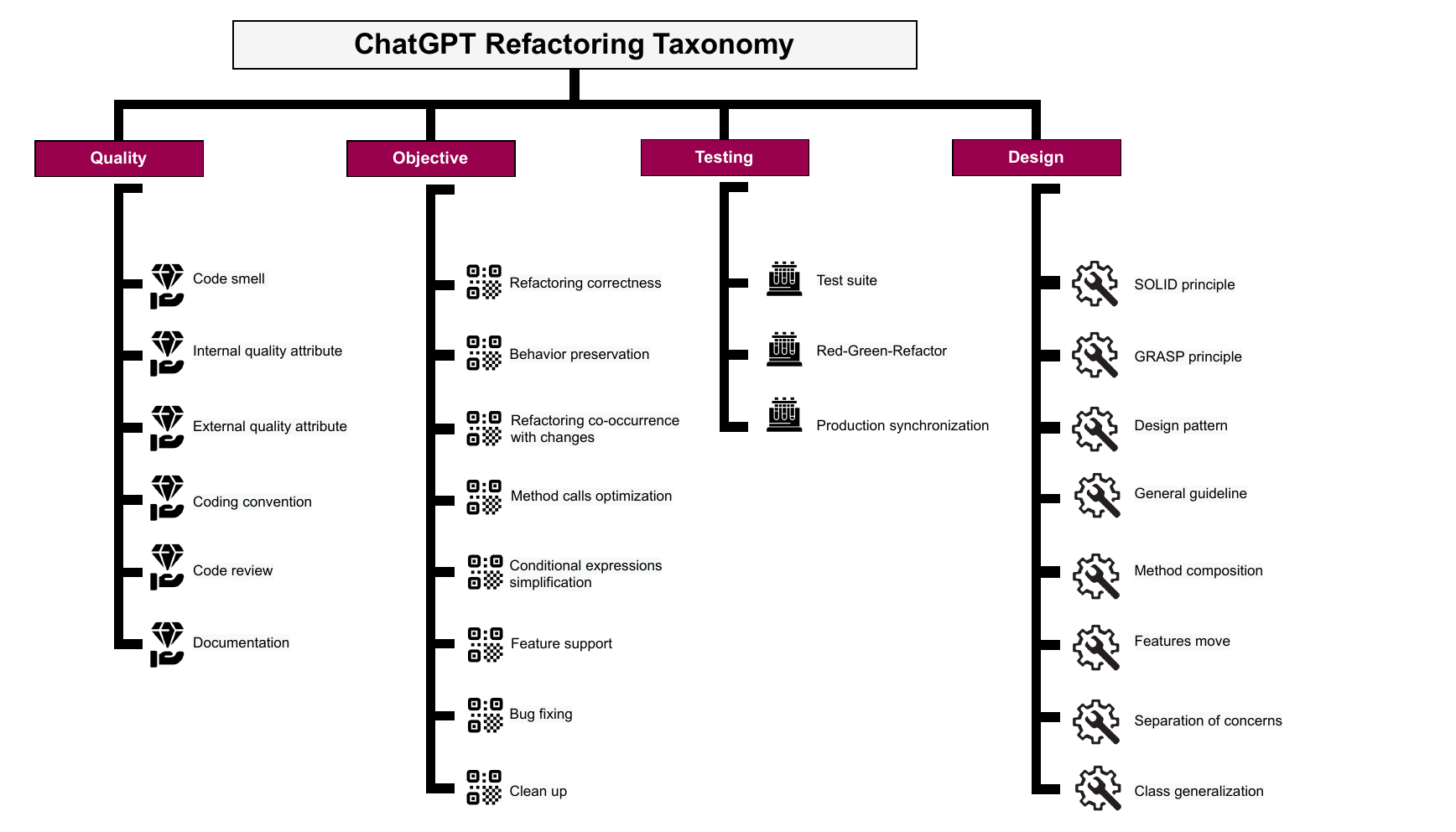}
\caption{\textcolor{black}{A taxonomy of the ChatGPT refactoring topics.}}
\label{fig:taxonomy}
\end{figure*}

\noindent \textcolor{black}{{After examining the refactoring responses provided by ChatGPT, we establish broad high-level classifications for the developer-ChatGPT refactoring dialogue.} Figure \ref{fig:taxonomy} shows
the taxonomy resulting from our analysis. The taxonomy is structured into two tiers: the upper tier includes four categories that organize activities with related objectives, while the lower tier comprises \taxonomy subcategories, offering detailed classification. Conversations regarding developer-ChatGPT refactoring revolve around four primary categories illustrated in the figure: (1) quality, (2) objective, (3) testing, and (4) design. It is important to highlight that our categorization is not exclusive; thus, a response might belong to multiple categories. 
 Examples for each category are listed in Table \ref{Table:example}. The ensuing part of this subsection delves deeper into these categories.}

\noindent\textbf{Category \#1: Quality.} When addressing refactoring requests, the quality of design plays a crucial role. Based on ChatGPT feedback, it ensures compliance with \textit{coding conventions}, promotes \textit{code reviews}, and maintains \textit{documentation} standards. Additionally, it aims to enhance both \textit{internal} and \textit{external quality attributes}, while steering clear of design pitfalls like \textit{code smells}. For example, ChatGPT advises on the best practices for code writing and the optimization of \textit{internal and external quality attributes}, as developers may not perceive the complete overview of the software design. 


\noindent\textbf{Category \#2: Objective.} This category compiles responses that emphasize assessing the accuracy of code transformations and determining if the proposed modifications result in a secure and reliable refactoring. These ChatGPT responses investigate topics such as \textit{refactoring correctness}, \textit{behavior preservation}, and the appearance of refactoring alongside other changes, in addition to responses related to refactoring operations such as \textit{method composition}, \textit{feature move}, and \textit{generalization}. Since developers frequently interleave refactoring with other activities, ChatGPT highlighted that combining refactoring with additional changes could obscure errors, thus increasing the risk of introducing bugs. Further, we have compiled instances where ChatGPT ultimately requests clear documentation of \textit{feature-related}, \textit{bug fix-related}, and \textit{clean up} tasks to enhance understanding of the rationale behind submitted refactoring requests. This demonstrates ChatGPT's recurring suggestions for improving developers' documentation practices, particularly concerning the perception and reasoning of the changes. 

\begin{table*}[htbp]
  \centering
	 \caption{\textcolor{black}{Examples of the ChatGPT refactoring topics.}} 
	 \label{Table:example}
\begin{adjustbox}{width=1\textwidth,center}
\begin{tabular}{llLllll}\hline
\toprule
\bfseries Category & \bfseries Sub-category & \bfseries Example (Excerpts from a related refactoring conversations) \\
\midrule
\multirow{8}{*} {\textbf{Quality}} & \cellcolor{gray!30}Code smell & \cellcolor{gray!30} \say{\textit{This version of the shell script refactors the existing solution to remove the duplication of the descriptor filename.}} 
\\ 
& Internal quality attribute & \say{\textit{Currently, many classes are tightly coupled with their dependencies. Try to inject dependencies instead of creating them inside the classes. This would make your classes easier to test and maintain.}} 
\\
& \cellcolor{gray!30}External quality attribute & \cellcolor{gray!30} \say{\textit{Your TopNav.js component is quite modular and you've done a good job organizing the related components. Below, I'll offer some refinements to increase the maintainability and readability of the code.}} 
\\
& Coding convention & \say{\textit{Consistent Naming Conventions:
Stick to one naming convention throughout your project for variables, methods, classes etc.}} 
\\
& \cellcolor{gray!30}Code review& \cellcolor{gray!30} \say{\textit{Review your refactored code and validate your use of inheritance.}} 
\\  
& Documentation&  \say{\textit{when creating a git commit the message is crucial for readability and understanding the purpose of the commit for future reference heres how you could craft a commit message for this refactoring.}} 
\\  
\hline
\multirow{18}{*} {\textbf{Objective}} & 
\cellcolor{gray!30}Refactoring correctness& \cellcolor{gray!30}\say{\textit{it's important to be mindful of Liskov's Substitution Principle (LSP), which states that objects of a superclass shall be able to be replaced with objects of a subclass without affecting the correctness of the program.}} 
\\ 
& Behavior preservation & \say{\textit{The EOF marker is quoted to prevent parameter substitution within the heredoc, preserving the exact content we want for the App.jsx file.}} 
\\
& \cellcolor{gray!30}\begin{tabular}[c]{@{}l@{}}\cellcolor{gray!30}Refactoring co-occurrences \\ \cellcolor{gray!30}with changes \end{tabular}  & \cellcolor{gray!30}\say{\textit{For brevity, I'm going to write a simple class demonstrating how you might implement some of these features using Google's JavaScript client library. Keep in mind this is just a starting point and you would need to expand upon this with proper error handling, user interface components, and more.}} 
\\
& Method calls optimization &  \say{\textit{Thank you for pointing out those mistakes i apologize for the confusion youre correct if the project is already configured to support es the r esm flag should not be needed and the temp file should be cleaned up heres the corrected \texttt{scriptcodeblockthis} corrected script should take care of the renaming refactoring and the update to \texttt{packagejson} without any lingering temp files.}} 
\\
& \cellcolor{gray!30}Conditional expressions simplification & \cellcolor{gray!30}\say{\textit{Let's refactor the Heading component to be a functional component. This refactor will also simplify the conditional logic and improve readability.}} 
\\
& Feature support & \say{\textit{You'd like to refactor your application's image caching system to use Google Drive instead of your backend server. We'll need to interface with Google Drive's API, and we will need to implement features such as saving image payloads, image files, pagination, pre-signed URLs, and deletion functionality.}} 
\\ 
& \cellcolor{gray!30}Bug fixing & \cellcolor{gray!30}\say{\textit{You are throwing exceptions but not catching them. Make sure to handle exceptions properly.}} 
\\
& Clean up & \say{\textit{The task involves refactoring and reorganizing the existing code into separate modules and folders for better manageability and separation of concerns.}} 
\\ 
\hline
\multirow{5}{*} {\textbf{Testing}} & \cellcolor{gray!30}Test suite & \cellcolor{gray!30}\say{\textit{Testing: You're using Cypress, Enzyme, and Jest, which is great for testing. Make sure the tests are updated as you refactor.}} 
\\
& Red-Green-Refactor &  \say{\textit{Red-Green Refactor is an excellent approach here. Start with a failing test (Red), implement your function, make the test pass (Green), and then refactor if necessary.}} 
\\
& \cellcolor{gray!30}Production synchronization & \cellcolor{gray!30}\say{\textit{Given the refactoring of the Game.js file into a new User.js file, you'll need to refactor the unit tests accordingly.}} 
\\
\hline
\multirow{20}{*} {\textbf{Design}} & SOLID principle & 
 \say{\textit{Some classes like \texttt{ContractRpcWriter} do a lot of work - initialization of Web3, gas estimation, transaction signing, and sending. Try to keep a class to a single responsibility. This will increase the modularity and readability of your code.}} 
 \\    
& \cellcolor{gray!30}GRASP principle &  \cellcolor{gray!30}\say{\textit{This design enables you to use polymorphism, one of the core principles in OOP, where a class has many ("poly") forms ("morphs").}} 
\\ 
& Design pattern &  \say{\textit{You're currently using the Strategy pattern in your design, where the Algorithm class defines a common interface for all strategies (algorithms), and Minimax and UserInput are concrete strategies that implement this interface. This pattern provides a good solution for your current requirements.}} 
\\ 
&\cellcolor{gray!30}General guideline & \cellcolor{gray!30}\say{\textit{Your refactored code looks much cleaner and modular! Dividing responsibilities between different classes is a good software engineering practice. It simplifies the understanding of code and makes maintenance easier.}} 
\\
& Method composition &  \say{\textit{This script will refactor the code by splitting it into separate files. It creates a new file \texttt{setupRoutes.js} to handle all the routing logic, deletes the handlers.js file as it's no longer needed, and moves the \texttt{generateHandler} function to a new file \texttt{generateHandler.js}.}} 
\\ 
& \cellcolor{gray!30}Features move &  \cellcolor{gray!30}\say{\textit{Creates a new directory named backend inside the src directory. Moves \texttt{server.js} and \texttt{servePromptDescriptor.js} to the backend directory. Refactors \texttt{server.js} by extracting the route handlers into a new file called \texttt{handlers.js}.}} 
\\
& Separation of concerns &  \say{\textit{The user class now encapsulates all the user-related functionalities. \texttt{getStoredUser}, \texttt{getPicks}, \texttt{updatePicks}, \texttt{backfillResults}, \texttt{switchUser}, and \texttt{displayBackfilledResults} functions are now methods of the User class. This provides a clear separation of user-specific logic and game logic.}} 
\\
& \cellcolor{gray!30}Class generalization & \cellcolor{gray!30}\say{\textit{Review your refactored code and validate your use of inheritance. Make sure that it adheres to Liskov's Substitution Principle (LSP). LSP is a concept in object-oriented programming that states that if a program is using a base class, it should be able to use any of its subclasses without the program knowing it.}} 
\\
\bottomrule
\end{tabular}
\end{adjustbox}
\vspace{-.3cm}
\end{table*}




\noindent\textbf{Category \#3: Testing.} \textcolor{black}{The goal of refactoring is to enhance internal software structure while keeping its functionality intact. Ideally, pre-existing unit tests should be adequate to confirm this consistency. However, since refactoring is often mixed with other tasks, alterations in software behavior might occur. In these scenarios, unit tests might fail to notice these changes unless they are updated to accommodate the new features. This issue has been highlighted in several ChatGPT responses, especially when developers overlook these behavior changes. Our analysis of these responses shows that ChatGPT suggests implementing unit tests prior to refactoring to ensure the code remains intact. It also recommends following the red-green-refactor methodology. Additionally, when developers seek guidance on testing during refactoring, ChatGPT recommends updating the test files. Thus, within the \textit{Testing} category, we identify the following sub-categories: \textit{test suite}, \textit{red-green-refactor}, and \textit{production synchronization}. 
}




\noindent\textbf{Category \#4: Design.} \textcolor{black}{According to ChatGPT responses, developers are asked to follow general guidelines, design pattern, and specific design principles (\eg SOLID and GRASP). ChatGPT's responses about the SOLID principle focus on key design principles such as the single responsibility principle, the Liskov substitution principle, and the dependency inversion principle. On the other hand, ChatGPT's responses about the GRASP principles offer guidelines for assigning responsibilities to classes such as low-coupling and high-cohesion. Additionally, it discussed refactoring changes such as \textit{method composition}, \textit{feature move}, and \textit{class generalization}. This indicates that ChatGPT asked developers to strive to design classes that are focused and easy to maintain by breaking down complex systems into more manageable components. 
}

\subsection{RQ$_2$: \textcolor{black}{In which scenarios does ChatGPT provide unsatisfactory responses during the refactoring conversations?}}

\noindent \textcolor{black}{In the refactoring process, ChatGPT's expressions of regret (such as admitting errors) and confirmations (such as maintaining confidence or agreement) are important. Regrets may be offered if the refactoring leads to errors or unexpected results. Affirmations serve to recognize refactoring successes.  Upon analyzing ChatGPT's responses, we cluster the responses into categories. 
}




\noindent\textbf{Category \#1: Apology.}  \textcolor{black}{When responding to errors in its refactoring suggestions, ChatGPT often employs phrases like `My apologies' and `I apologize.' These expressions serve to acknowledge mistakes and reflect a promise to rectify them. ChatGPT expresses regret when it introduces bugs in the code, misinterprets the code's content, overlooks update capabilities, incorporates poor practices, or displays misunderstanding in refactoring. ChatGPT response exemplifies this category is shown in Figure \ref{apology-affirmation}. As can be seen, ChatGPT apologizes for introducing duplicate code, which is considered one of the bad practices that violates the best design principles. This demonstrates ChatGPT's awareness of the importance of maintaining code quality and adhering to established coding standards. Next,  we provide an analysis of these categories:}
\begin{figure}
\begin{boxA}
\underline{\textbf{Example of apology \cite{apology}:}}
\\ 
You're correct! My apologies for the redundancy in the code. We can indeed refactor the `\textbf{TasksList}' component to use the `\textbf{PromptDescriptorViewer}' component instead of re-implementing its functionality. This would allow us to uphold the DRY (Don't Repeat Yourself) principle in our code.
\end{boxA}
\begin{boxA}
\underline{\textbf{Example of affirmation \cite{affirmation}:}}
\\ 
Certainly, it's a good idea to create a separate function for fetching the descriptor URL. We can place this function in a common utilities or services file. \\
Let's refactor the code to make this change. We will create a new file `\textbf{apiServices.js}' for such service functions. 
\end{boxA}
\caption{\textcolor{black}{Chatbot interaction in the context of refactoring.}}
\label{apology-affirmation}
\vspace{-.4cm}
\end{figure}
\begin{itemize}
    \item \textbf{Bugs introduction.} ChatGPT apologizes when its refactoring suggestions introduce bugs or errors into the code. For example, it may suggest a change that inadvertently breaks the functionality of the program. This type of apology is crucial as it reflects the model's recognition of its mistake and its impact on the code’s functionality.
    \item \textbf{Lack of understanding.} ChatGPT sometimes misinterprets the developer’s code or intent. It apologizes for any misunderstandings about refactoring and tries to clarify or correct its previous suggestions.
    \item \textbf{Missing update functionality.} When ChatGPT fails to account for necessary updates across different parts of the code, it issues an apology. This demonstrates the need for comprehensive changes that span multiple areas of the codebase.
    \item \textbf{Bad practice introduction.} ChatGPT’s suggestions may violate coding best practices. In such cases, it acknowledges the mistake and attempts to rectify it.
\end{itemize}

\noindent\textbf{Category \#2: Affirmation.}  \textcolor{black}{In its responses, ChatGPT includes keywords like `certainly', `absolutely', `of course', and `sure' to affirm the developer's prompt on refactoring. These words reflect a high level of agreement or assurance in addressing the developer's question or suggestion. ChatGPT consistently expresses this degree of confidence or agreement throughout its refactoring suggestions. When recommending refactoring, ChatGPT aims to improve code quality, offers examples for code refactoring, adheres to established refactoring standards or principles, and follows best coding practices, such as modularization and abstraction, to boost code maintainability and readability. Additionally, it introduces alternative refactoring methods to give developers a range of strategies or techniques to explore. Figure \ref{apology-affirmation} presents an example of the ChatGPT responses that exemplify this category. As shown, ChatGPT adheres to the best practice of creating separate functions with common logic. Below, we provide an analysis of these categories:}

\begin{itemize}
    \item \textbf{Code quality improvement.} ChatGPT affirms its goal of improving code quality through its suggestions. This includes applying best coding practices to enhance maintainability and readability.
    \item \textbf{Refactoring demonstration.} When affirming a refactoring suggestion, ChatGPT provides concrete examples and guidance, demonstrating how to implement the refactoring effectively.
    \item \textbf{Code guideline adherence.} ChatGPT affirms its understanding of the developer’s intent by providing guidelines. This reassurance helps in building trust and ensures that the developer feels understood.
    \item \textbf{Best practice application.} ChatGPT often affirms its adherence to best practices in its suggestions. This is important for ensuring that the code remains maintainable and follows coding standards.
    \item \textbf{Refactoring extension.} In addition to affirming a single refactoring suggestion, ChatGPT sometimes suggest applying additional refactorings.
\end{itemize}



In the course of this analysis, we identified several potential problems with the responses, which were not directly related to the refactoring request. In fact, we witnessed several instances of issues encompassing reliance on outdated information, technical errors, incorrect API calls, and fabricated source classes. Although these mistakes can have significant impacts, we are not evaluating their severity in this study, as our focus is on the refactoring context. Investigating the severity of these errors is an intriguing subject that would necessitate more information, such as the complete source code of the projects. However, such data is not available in the current dataset we utilize. So, as part of future work, we plan to gather such relevant information and conduct a severity analysis since it would be interesting to investigate whether refactoring would increase the probability of these mistakes.

\subsection{RQ$_3$: \textcolor{black}{What is the performance of the proposed prompt, compared to the dataset?}}



\begin{figure}
\begin{boxA}
\textit{\textcolor{blue}{\textbf{Note for developer:} text written in blue is meant for your eyes only, not as prompt for chat. Items in \textless \textgreater    should be filled in as your own texts, not to keep.}}

\textit{\textcolor{blue}{\textbf{Prompting techniques:} few shots, one shot, zero shot, context constraints, output constraints, chain of thought.}}


You are an expert in programming, refactoring, and providing advice in software quality. Given the following prompt/code snippet, modify/write the code so that it best matches what the user wants, and provide comments explaining your work. If needed, ask clarifying questions to check both your and the user`s understanding of the prompt. 

    \textbf{\textcolor{black}{Working Set:}} 

    \textit{\textcolor{blue}{\textless here you will input in the code you are working with. Make sure to state clearly if they are form the same file or different files to avoid confusion.\textgreater}}
    
    \textbf{\textcolor{black}{Context:}}
    
    - Language: \textit{\textcolor{blue}{\textless e.g., JavaScript, Python, Java\textgreater  }}
    
    - Project: \textit{\textcolor{blue}{\textless brief description of how/where this code is used\textgreater  }}
    
    - Installed Tools:  \textit{\textcolor{blue}{\textless list dependencies, libraries, frameworks\textgreater  }} 
    
    - Constraints:  \textit{\textcolor{blue}{\textless e.g., Do not change the output behavior, Maintain API structure, output constraints\textgreater  }} 
    
   \textbf{\textcolor{black}{Refactoring Task:}}
   
   \textit{\textcolor{blue}{\textless give refactoring prompt as much detail as possible, give two - three full sentences MINIMUM. Use these keywords as a starting context, these are meant to be used in the beginning of the action:}}
   
           \textit{\textcolor{blue}{Refactoring Patterns (Self-Affirmed Refactoring)}}
           
   \textit{\textcolor{blue}{\textless refactoring pattern-related quality issues words such as these, after the initial action. Start with ``to improve...'':\textgreater  }}
           \textit{\textcolor{blue}{Refactoring Intent (Self-Affirmed Refactoring)}}

   -  \textit{\textcolor{blue}{\textless explicitly state any components that you want/would prefer chat to use.\textgreater  }}

   \textit{\textcolor{blue}{\textless To improve the code:\textgreater  }}

   - \textit{\textcolor{blue}{\textless Clearly state any components or patterns that should be used.\textgreater  }}

   - \textit{\textcolor{blue}{\textless Ensure modularization where applicable.\textgreater  }}

   - \textit{\textcolor{blue}{\textless Follow established design patterns.\textgreater  }}

   \textbf{\textcolor{black}{Steps to Follow:}} 
   
   1. Analyze the Code: Identify areas that require refactoring.
   
   2. Apply Refactoring Strategies: Improve quality attributes (e.g., readability, maintainability, efficiency, performance, cohesion, coupling, etc). 
   
   3. Provide Multiple Solutions: Generate at least two refactored versions with explanations. 
   
   4. Validate Functionality: Ensure the new code behaves the same as the original. 
   
   5. Comment \& Document: Provide commit messages and inline comments explaining the changes. 
   
  \textbf{\textcolor{black}{Output Format:}}
  
  - Provide refactored code inside a code block.
  
  - Include a before-and-after comparison.
  
  - Provide concise commit messages summarizing each major change.
  
  - Generate test cases to validate the refactoring. 

   \textbf{\textcolor{black}{Example start:}}
   
   \textit{\textcolor{blue}{\textless how it is being ran\textgreater  }} 
   
   \textbf{\textcolor{black}{Example end}}

  If there are any unclear instructions, please ask clarifying questions.

\end{boxA}
\caption{\textcolor{black}{Refactoring prompt template.}}
\label{template}
\end{figure}


\noindent Prompt Engineering is a technique for guiding LLMs to generate specific outputs. The need for studying prompting in LLM-driven code refactoring comes from the need to improve its effectiveness, and reliability. While LLMs have shown promise in automating refactoring tasks \cite{gehring2023deterministic}, their performance is highly dependent on how prompts are structured and syntactically formed. Through studies, different refactoring tasks require varying levels of context and reasoning. Simpler refactorings like renaming can often be handled with basic prompts. However, more complex refactorings, such as the extractions, demand more structure, iterative, and context-based prompts to ensure correctness. 

\textcolor{black}{Our goal is to guide developers by creating refactoring prompt template that can reduce time and effort, allowing the model to adapt to a variety of refactoring tasks. Figure \ref{template} illustrates our proposed prompt template for conversational refactoring using LLMs. The crafted template consists of the following main sections:}
\begin{itemize}
\item \textcolor{black}{\textbf{Role.} The prompt starts by specifying a role as an AI expert in refactoring and software quality. Reynolds and McDonell mentioned that persona-based pattern (\ie AI is assigned a specific role), significantly improves contextual understanding \cite{PPforLLM}.}
    \item \textcolor{black}{\textbf{Working Set.} This section allows developers to provide their code for refactoring purposes.}
    \item \textcolor{black}{\textbf{Context.} This section allows developers to indicate the context pertaining to the language, project, installed tools, and constraints.}
    \item \textcolor{black}{\textbf{Refactoring Task.} This section is the main part that allows developers to provide specific refactoring tasks, including motivation and intended impact \cite{di2018preliminary}.  
    Since recent study has shown that developers often make generic refactoring requests, while ChatGPT typically includes the refactoring intention \cite{alomar2024refactor}, our goal is to assist developers in utilizing the right information in the prompt so that LLMs can focus on the intended improvement instead of making random changes.}
    \item \textcolor{black}{\textbf{Steps to Follow.} This section ensures a step by step instruction of how LLMs should proceed with refactoring task, while ensure behaviour preservation and successful test results.}
    \item \textcolor{black}{\textbf{Output Format.} This section specifies the output format of the refactored code requested by the developer.}
     \item \textcolor{black}{\textbf{Example.} This section specifies the example provided by the developer.}
    \item \textcolor{black}{\textbf{Clarification Option.} This part ensures LLMs understanding of the information given in the prompt.}
\end{itemize}
\begin{table}
\centering
\caption{\textcolor{black}{Statistics of LLM prompt efforts.}}
\label{Table:RQ3_results}

\begin{adjustbox}{width=0.5\textwidth,center}
\begin{tabular}{@{}lllllll|llllll|ll@{}}
\toprule
\multicolumn{1}{c}{\multirow{3}{*}{\textbf{Metrics}}} & \multicolumn{6}{c|}{\textit{\textbf{Developer's prompt}}} & \multicolumn{6}{c|}{\textit{\textbf{Our prompt}}} & \multicolumn{2}{c}{\textit{\textbf{Statistical difference}}}  \\ \cmidrule(l){2-15} 
\multicolumn{1}{c}{} & \multicolumn{1}{c}{\textbf{Min}} & \multicolumn{1}{c}{\textbf{Q1}} & \multicolumn{1}{c}{\textbf{Median}} & \multicolumn{1}{c}{\textbf{Mean}} & \multicolumn{1}{c}{\textbf{Q3}} & \multicolumn{1}{c|}{\textbf{Max}} & \multicolumn{1}{c}{\textbf{Min}} & \multicolumn{1}{c}{\textbf{Q1}} & \multicolumn{1}{c}{\textbf{Median}} & \multicolumn{1}{c}{\textbf{Mean}} & \multicolumn{1}{c}{\textbf{Q3}} & \multicolumn{1}{l|}{\textbf{Max}} &
\multicolumn{1}{c}{\textbf{\textit{p}-value}} & \multicolumn{1}{l}{\textbf{Cliff's delta ($\delta$)}}\\ 
\midrule
\cellcolor{gray!30}Number of turns & \cellcolor{gray!30}1 & \cellcolor{gray!30}2 & \cellcolor{gray!30}3&  \cellcolor{gray!30}13.58 & \cellcolor{gray!30}9.50 & \cellcolor{gray!30}11 & \cellcolor{gray!30}1 &\cellcolor{gray!30}1 &\cellcolor{gray!30}1 & \cellcolor{gray!30}1.45  &\cellcolor{gray!30}2  & \cellcolor{gray!30}3 &  \cellcolor{gray!30}\textbf{0.000005649} & \cellcolor{gray!30}large (0.56)\\
Prompt length & 372& 2996 & 4017 & 16604 & 7812 & 13537 &  1603& 3530 & 4489 & 4855.52  & 5287.50 & 6570 & 0.1952 & small (0.04) \\
\cellcolor{gray!30}Response length & \cellcolor{gray!30}1497 & \cellcolor{gray!30}3865 & \cellcolor{gray!30}5871 & \cellcolor{gray!30}27538.23 & \cellcolor{gray!30}21049.50 & \cellcolor{gray!30}23043& \cellcolor{gray!30}2297 & \cellcolor{gray!30}4192 & \cellcolor{gray!30}5778 & \cellcolor{gray!30}6975.71 & \cellcolor{gray!30}7630 & \cellcolor{gray!30}11445 & \cellcolor{gray!30}0.6376 & \cellcolor{gray!30}small (0.11) \\
\bottomrule
\end{tabular}
\end{adjustbox}
\end{table}
\textcolor{black}{Regarding the evaluation of the proposed prompts, we run our prompt on the same code provided by the original DevGPT developers and calculate the three metrics. Looking at Table \ref{Table:RQ3_results}, we found
that our prompts differ (\ie fewer turns ($\mu$ = 1.45), shorter prompt ($\mu$ = 4855.52), and shorter response ($\mu$ = 6975.71))  from original developer's prompts. We also performed a
non-parametric Mann-Whitney U test and we obtained a statistically significant \textit{p}-value when the values of these two groups were
compared (\textit{p}-value \textless 0.05 for number of turns), and accompanied
with a small or large effect size depending on the prompt
effort/metric.  We notice that usually when the developer provides the information listed in our prompt, the refactoring conversations are shorter and more fruitful.}

\section{Discussion}
\label{Section:Discussion}

\noindent\textbf{ \textcolor{black}{Takeaway \#1:  \textit{ Effective prompting of ChatGPT for minimal-interaction refactoring responses remains an area for further exploration.}}} \textcolor{black}{Since our study aims to constitute a \say{successful} prompt in the context of refactoring, we propose a structured prompt template to help developers get a satisfactory response from GPT with the least possible shots (\ie prompt rounds). This effective structured prompt pattern serves as a guideline for prompting related to refactoring and may have the potential to be very useful to practitioners who use GPT for refactoring tasks. For example, it can improve accuracy, relevance, developer productivity, and save time and effort (RQ$_3$).}

\noindent\textbf{ \textcolor{black}{Takeaway \#2:  \textit{ LLM-Driven refactoring is sub-optimal for certain refactoring contexts.}}} \textcolor{black}{Our taxonomy shows that developers prompted LLMs for various refactoring operations, ranging from method composition, feature move, data organization, simplifying method calls, to dealing with generalization. This sheds light on exploring two important directions: (1) what refactoring operations that LLMs can/cannot handle. This was observed in the apologetic and affirmation phrases due to the challenge of LLMs can only maintain a limited understanding of the broader context (RQ$_2$), and (2) what refactoring operations are considered underrepresented (RQ$_1$). 
 \textcolor{black}{For apologetic cases, we found that ChatGPT struggled the most with situations involving bug introduction, lack of understanding, and the introduction of bad practices. In the DevGPT example \cite{formatting}, the developer had to prompt GPT 11 times to obtain a satisfactory response, with ChatGPT repeatedly apologizing for misunderstandings, oversights, and confusion. 
 Consequently, we believe that developers can change their prompting strategy to get a satisfactory result with fewer turns by improving prompts with clarity, requesting GPT to generate multiple solutions, and asking for reasoning/explanation behind the solutions. These strategies have been incorporated into our prompt template, which acts as a guide for developers when performing conversational refactoring.}}

\noindent\textbf{ \textcolor{black}{Takeaway \#3:  \textit{ LLM-generated code can be unreliable; requiring strict processes and methods to detect and highlight unreliable, generated code.}}} \textcolor{black}{According to our analysis, LLMs can provide correct refactoring-related responses. However, there are cases where LLMs can hallucinate or suggest sub-optimal solutions, which especially shown in Developer-ChatGPT refactoring conversation (RQ$_1$).} \textcolor{black}{According to  RQ$_3$, while compared to the developer prompts, we had fewer iterations of prompting needed, there were still certain issues that come from automating refactoring using ChatGPT. Issues that have been mentioned by previous study \cite{chen2024chatgpt}, such as ChatGPT-4's abilities in adhering to formatting based off the requirements set by the developers. For example, in multiple of our tests, when it involves ChatGPT to write markdown files, such as README.mds \cite{formatting}, ChatGPT rarely succeeds in doing so. We analyze the failures that came from such behaviors and find that it is likely that ChatGPT is unaware of when and where it has left a markdown code box. This leads it to create awkward instances of responses where information that should be retained in markdown jumps out of the markdown block, and instances where ChatGPT was adding an afterword in its response but ended up in a code block.}

\textcolor{black}{Outside of code formatting, we also noticed that in instances where we ask ChatGPT to provide commit messages, it often fails to do so in Git format (\eg git commit -m ``Hello World!''). This could mean various issues, such as ChatGPT not understanding the needs of the developers or having formatting issues again. ChatGPT often fails to follow the entire prompt. One of the common mistakes made by ChatGPT was skipping creating test cases and creating multiple versions of the code and assessing them by comparisons. We had made those two requirements a stable in our prompt formatting, but most of the time when tested ChatGPT either completely ignores it or asks us to prompt again if we want the test cases. Additionally, when it comes to ChatGPT’s refactoring suggestions, it has the potential to hallucinate. For instance, when applied to real-world Extract Method instances, ChatGPT-3.5 was found to produce two types of hallucinations: providing refactoring suggestions with either syntactical errors or illogical code that does not address the issue, such as suggesting to extract an entire method body. When reviewing the original developer prompts from DevGPT, we found that ChatGPT sometimes produced illogical code when asked to produce markdown files, and it exhibited another hallucination where it misunderstood the location of a file within a project’s file structure, both causing developers to make additional prompts to address these issues. Developers can take steps to minimize these hallucinations by using tools that filter out incorrect or impractical suggestions. For example, Pomian \etal \cite{pomian2024together} developed \textsc{EM-Assist} that uses LLMs to generate, validate, enhance, and rank suggestions. The \textsc{EM-Assist} IntelliJ plug-in generates 5-10 suggestions for the selected method, employing program slicing (\eg excluding code that does not affect a target variable or statement) to enhance the generated suggestions, then using the IntelliJ IDEA API to filter invalid suggestions by verifying whether the code meets the refactoring preconditions for Extract Method. Finally, the plug-in ranks the suggestions based on how frequent a particular refactoring is suggested by the LLM, and presents them to the developer to select from. Thus, developers can use tools such as \textsc{EM-Assist} to reduce the amount of suggestions with hallucinations, while also spending less time finding the most useful suggestions for their issue. Ultimately, further research exploring and addressing these types of hallucinations is needed to increase the reliability of ChatGPT’s suggestions and LLM-based refactoring in general.}

\section{Threats To Validity}
\label{Section:Threats}

\noindent\textbf{External Validity.} \textcolor{black}{We center our study on open-source systems. Consequently, our findings may not be generalized to all other open-source systems or commercially developed projects. Moreover, although our dataset covers different programming languages, our results may not generalize to systems written in other languages that are not considered in this study. Nevertheless, the objective of this paper is not to construct a theory universally applicable to all systems, but to demonstrate the capability of ChatGPT to provide solutions for code refactoring. Another potential concern pertains to the proposed taxonomy. It may not apply universally to other open-source or commercial projects. Therefore, we cannot claim that the findings regarding ChatGPT refactoring taxonomy are applicable to all software systems, especially those where the imperative for design enhancement might be less significant. Finally, our study is solely focused on ChatGPT, and it is recommended that future research extend to other general-purpose chatbots applicable to software development, such as the recently released Google Bard, which may have limitations that ChatGPT does not possess \cite{tufano2024unveiling}.}

\noindent\textbf{Internal and Construct Validity.} \textcolor{black}{Concerning the identification of developer-ChatGPT refactoring conversations, we constructed our dataset by extracting data source types that contain the term `refactor' in the prompt. There is the possibility that we have excluded synonymous terms/phrases. However, even though this approach reduces the number of source types in our dataset, it also decreases false positives, and ensures that we analyze artifacts that are explicitly focused on refactorings. Another potential concern regarding validity involves the developer-ChatGPT refactoring conversation. Given that refactorings may be interleaved with other modifications (\ie developers executed changes alongside refactorings) \cite{murphy2012we}, we cannot assert that the chosen refactoring conversations solely pertain to refactoring activities. Moreover, 
 throughout the manual analysis, we intentionally omitted cases where ChatGPT's contribution to refactoring activity was ambiguous, and we applied multiple labels in cases where ChatGPT served multiple purposes. \textcolor{black}{Finally, we designed our refactoring prompt template with generalizability in mind but tested it with GPT to be consistent with DevGPT dataset, its effectiveness still need to be assessed using other LLMs.}}

\section{Conclusions}
\label{Section:Conclusions}

\textcolor{black}{Large Language Models (LLMs) have seen a rise in popularity and are extensively utilized in numerous software engineering tasks. These tasks include, but are not limited to, refactoring, testing, code review, and program comprehension. This paper aims to explore interactions between developers and ChatGPT concerning refactoring, aiming to gain insight into how developers identify areas for code improvement and how ChatGPT addresses their needs. Our approach consists of extracting refactoring-related discussions from 29,778 ChatGPT prompts and responses, with an emphasis on developers' clear refactoring objectives. Our results show that 
(1) developer-ChatGPT refactoring interactions cover \taxonomy themes,  (2) ChatGPT’s affirmations and apologies in multi-turn conversations highlight failure points, helping us study when and how it falls short, and (3) our refactoring prompt template helps developers receive precise, efficient, and satisfactory responses with the least amount of interaction.  
}

\noindent{\textbf{Declaration of generative AI and AI-assisted technologies in the writing process.}}
 During the preparation of this work, the author used the ChatGPT Web interface and Writefull to improve the language and readability of some sections. After using this tool, the authors reviewed and edited the content as needed and take full responsibility for the content of the publication.

\bibliography{sample-base}
\bibliographystyle{ieeetr}

\end{document}